\documentstyle[prb,aps_mod,twocolumn
]{revtex_mod}

\begin{document}

\title{Submonolayer epitaxy with impurities}
\author{Miroslav Kotrla}
\address{Institute of Physics, Czech Academy of Sciences, Na Slovance 2,
182~21 Praha 8, Czech Republic}
\author{Joachim Krug}
\address{Fachbereich Physik, Universit\"{a}t GH Essen, 45117 Essen,
Germany}
\author{Pavel \v{S}milauer}
\address{Institute of Physics, Czech Academy of
Sciences, Cukrovarnick\'a 10, 162~53 Praha 6, Czech Republic}

\date{\tt Submitted to Physical Review B on January 25, 2000}

\abstr{
\begin{abstract}
The effect of impurities on epitaxial growth in the submonolayer
regime is studied using kinetic Monte Carlo simulations
of a two-species solid-on-solid
growth model. Both species are mobile, and attractive interactions
among adatoms and between adatoms and impurities are incorporated. 
Impurities can be codeposited with the growing material or predeposited
prior to growth. The activated exchange of impurities and adatoms is
identified as the key kinetic process in the formation of a growth morphology
in which the impurities decorate the island edges. The dependence of 
the island density on flux and coverage is studied in detail. 
The impurities strongly increase the island density without appreciably
changing its power-law dependence on flux, apart from a saturation of
the flux dependence at high fluxes and low coverages. A simple analytic
theory taking into account only the dependence of the adatom diffusion constant
on impurity coverage is shown to provide semi-quantitative agreement with
many features observed in the simulations. 
\end{abstract}
}

\maketitle

\section{Introduction}

Recent progress in the fabrication of atomically smooth interfaces
by Molecular Beam Epitaxy (MBE) has lead to an increasing
appreciation of the dramatic, detrimental or beneficial, effects
that small amounts of impurities may have on the morphology of
growing films. Adsorbates acting as {\em surfactants\/} can stabilize
layer-by-layer growth of metal 
\cite{vegt92,rosenfeld93,vrijmoeth94,scheffler95,wulfhekel96,meyer96}
and semiconductor \cite{voigtlaender95,kandel99} surfaces. On the other hand,
for the simple case of Pt(111) homoepitaxy it was recently 
shown \cite{kalff98} 
that minute coverages of CO strongly increase the step edge barriers for
interlayer transport, thus enhancing three-dimensional mound 
growth.\cite{kalff99} The effect of additional surface species on 
growth and nucleation is of obvious importance also in more complex,
technologically relevant deposition techniques such as chemical
vapor deposition.\cite{andersohn96}
In either case the detailed atomistic
kinetics and energetics of the interaction between adsorbate and deposited 
material influence the growth mode to a degree which makes it very
difficult to formulate general rules for large classes of growth systems.

As a first step towards an improved understanding of the {\em generic\/} 
effects of impurities on epitaxial growth, in the present paper we introduce
a minimal model which, we hope, is simple enough to extract some insights
of fairly general validity, and yet possesses sufficient flexibility to
include most physically relevant microscopic features. The model is based
on the standard solid-on-solid description for the growth of a simple
cubic crystal.\cite{weeks79,kotrla97} The impurities are represented
by a second particle species, which can be codeposited with the
growing material or predeposited prior to growth. Impurities diffuse and
interact attractively with the deposit atoms (adatoms), 
but they do not attract
each other and hence do not nucleate islands. The details of the model
are described in Section~\ref{Model}. A brief account of some preliminary
results was given in an earlier communication.\cite{ecoss18}  

Despite its simplicity, the model contains a large number of parameters:
impurity and growth fluxes, substrate temperature, and energy barriers 
for half a dozen
kinetic processes. To focus our efforts, we concentrate on modeling
a situation in which the impurities {\em decorate\/} the island 
edges, forming a monatomic chain along the island
perimeter. Preferential adsorption of impurities at step edges is
suggested by bond counting arguments, and has often been invoked to
explain the strong effect of submonolayer adsorbate coverages
on growth behavior, e.g., through a change of the barrier for
interlayer transport.\cite{vegt92,kalff98,markov94} 

It will be shown 
in Section~\ref{Morphology} that the growth of decorated islands 
requires,
in addition to a suitable choice of binding energies, the possibility
of impurity-adatom exchange. Such a process, which is a
two-dimensional analog of the exchange mechanism responsible
for the floating of surfactants in multilayer growth,\cite{meyer96} 
was recently demonstrated to play a crucial role in the submonolayer 
homoepitaxy of Si(001) in the presence of hydrogen.\cite{smilauer98}    

In Section~\ref{Island} the influence of the adsorbates on the island
density is investigated, both for codeposited and predeposited
impurities. In the absence of impurities the scaling relation 
\begin{equation}
\label{N}
N \sim (F/D)^\chi
\end{equation}
between island density $N$, deposition flux $F$ and adatom diffusion
coefficient $D$ is well established 
theoretically,\cite{stoyanov81,venables84,wolf95} numerically 
\cite{wolf95,tang93,evans93,ratsch94,amar95}
and experimentally.\cite{stroscio94}  In two dimensions, 
rate equation analysis 
\cite{stoyanov81,venables84,wolf95,tang93,evans93} yields the expression
\begin{equation}
\label{gamma}
\chi = \frac{i^\ast}{i^\ast+2}
\end{equation}
for the exponent $\chi$ in terms of the 
size $i^\ast$ of the largest unstable cluster. 

There are several 
conceivable mechanisms by which attractive 
impurities \cite{liu95} could alter the relationship
(\ref{N}). First, impurities may act as nucleation centers, thus
effectively {\em decreasing} $i^\ast$ and therefore $\chi$; in the
extreme case of immobile adatom traps the limit of spontaneous
nucleation with $i^\ast = \chi = 0$ would be 
realized.\cite{amar95,chambliss94} Second, impurities decorating the island
edges may induce energy barriers to attachment. Kandel \cite{kandel97}
has shown that, provided these barriers are sufficiently strong,
the exponent $\chi$ in (\ref{N}) is {\em increased\/} such that
(\ref{gamma}) is replaced by $\chi = 2 i^\ast/(i^\ast + 3)$.
Both mechanisms imply an 
increase of the island density compared to the case of pure
homoepitaxy.

Our simulations indicate that none of these two mechanisms 
are operative under the conditions used in our model: The addition
of impurities is found to increase the island density in all cases,
but the scaling of $N$ with the flux $F$ remains unaffected within the
accuracy of the simulation. An analysis of the relevant microscopic
processes \cite{ecoss18} reveals that, within our model, 
even completely decorated island edges do not provide efficient
barriers to attachment, and therefore the scenario of Kandel 
\cite{kandel97} does not apply. 

In view of (\ref{N}), an increase of the island density at fixed
$\chi$ suggests that the
main effect of the impurities is to reduce the mobility $D$ of adatoms. 
The reduction of the adatom mobility has been identified as the most
important mechanism contributing to the surfactant action of 
Sb on Ag.\cite{vrijmoeth94,scheffler95,meyer95} 
Decreasing $D$ reduces the island size and favors the growth of
ramified, rather than compact islands. Both effects enhance interlayer
transport, since the adatoms landing in the second layer have
more opportunities to descend, and thus promote layer-by-layer growth.
In our model the island size decreases but the island shapes remain
compact, because the edge decoration facilitates edge diffusion
(see Section~\ref{Morphology}).

For an analytic description of the relation between impurity
coverage, adatom mobility and island density, 
in Section~\ref{Rateq} we develop a simple rate equation approach
which provides a semi-quantitative explanation for many (though not all) 
features observed in the simulations. Some conclusions and open
questions are formulated in Section~\ref{Conclusions}.

\section{Model}
\label{Model}

The growth model employed in this work 
has been briefly described in an earlier paper.\cite{ecoss18}
It is a solid-on-solid model 
with two surface species $A$ and $B$, where 
$A$-particles correspond to the growing material, and  
$B$-particles represent the impurities. The
simulation starts on a flat substrate composed only of $A$-atoms.
The basic microscopic processes are deposition and migration;
desorption is not allowed. 
Two deposition modes are considered:
(i) simultaneous deposition ({\em codeposition\/}) of both species and
(ii) {\em predeposition\/} of a certain impurity coverage
prior to growth. In the case of codeposition 
the fluxes $F_A$ and $F_B$ of the two species may differ.

The migration of a surface atom is modeled as a nearest--neighbor hopping
process with the rate $R_D=k_0 \exp (-E_D /k_B T)$,
where $k_0= 10^{13}$ Hz is an adatom vibration frequency,
$E_D$ is the hopping barrier, $T$ is the substrate temperature
and $k_B$ is Boltzmann's constant.
The hopping barrier is the sum of a term from the substrate
$E_{\rm sub}$ and a contribution from each lateral nearest neighbor 
$E_{\rm n}$. Both
contributions depend on local composition: For each term we have the
four possibilities $AA$, $AB$, $BA$ and $BB$.
The hopping barrier of an atom $X$ (of type $A$ or $B$) is then
\begin{equation}
\label{E_D^X}
E_D^X= \sum_{Y=A,B} \left( n_0^Y E_{\rm sub}^{XY}
+ n_1^{XY} E_{\rm n} ^{XY} \right) ,
\end{equation}
where $E_{\rm sub}^{XY}$ is the hopping barrier for a free $X$ adatom 
on a substrate atom $Y$, $n_0^Y$ is equal to one if a substrate
atom is of type $Y$ and zero otherwise,
$n_1^{XY}$ is the number of nearest-neighbor
$X$-$Y$ pairs, and $E_{\rm n}^{XY}$
is the corresponding contribution to the barrier (symmetric in $X$ and $Y$).
Lateral interactions between impurity atoms are neglected
($E_{\rm n}^{BB}=0$).

In the simulations reported in this paper we used 
$E_{\rm sub}^{AA}=0.8$~eV,
$E_{\rm sub}^{AB}=0.1$~eV,
$E_{\rm sub}^{BA}=1.0$~eV,
$E_{\rm sub}^{BB}=0.1$~eV, 
and the substrate temperature $T=500$~K.
The low values of $E_{\rm sub}^{AB}$ and $E_{\rm sub}^{BB}$ ensure that
atoms deposited on top of an impurity instantaneously descend to the 
substrate.
Growth and impurity fluxes were varied 
in the interval ranging from 0.00025~ML/s to 0.25~ML/s.
The system sizes ranged from 300$\times$300 to 500$\times$500.
The nearest-neighbor coupling 
$E_{\rm n}^{AA}$ between $A$-atoms controls \cite{ratsch94} the 
size of the critical nucleus $i^\ast$.
It may be stronger ($E_{\rm n}^{AA}  > E_{\rm n}^{AB}$) or 
weaker ($E_{\rm n}^{AA}  < E_{\rm n}^{AB}$) than the coupling
to the impurities. 
In equilibrium at low temperatures,  
the former case leads to the
formation of islands composed inside mainly of $A$ atoms
with $B$ atoms bounded near the edges,
while in the latter case it is energetically more 
favorable when $B$ atoms are inside the islands.

However, our simulations show that  growth leads to intermixing
of $A$ and $B$ atoms
in both cases, $E_{\rm n}^{AA}  > E_{\rm n}^{AB}$ {\em and\/}
$E_{\rm n}^{AA}  < E_{\rm n}^{AB}$.
Thus the energetic bias 
favoring segregation is not sufficient to obtain
configurations with impurities mostly
at island edges (decorated islands). To achieve this, we have to 
introduce an additional thermally activated process, which allows
an $A$ atom approaching an island 
to exchange with an impurity covering the island edge.
A similar process was introduced previously \cite{smilauer98}
in the context of homoepitaxy on Si(001) with predeposited
hydrogen. In that work, an
$A$-atom was allowed to exchange with an impurity provided the $A$-atom
was not bonded to another $A$-atom at a nearest-neighbor site.
In our case
this modification turned out not to be sufficient,
since impurities were still 
found to be progressively trapped inside islands during
growth.
We therefore allow the exchange of an $A$ atom with an impurity also
when it has a {\em single\/} bond  to another $A$ atom
in a nearest-neighbor position. Using this rule, which is analogous
to the exchange process invoked in the case of three-dimensional 
growth with surfactants,\cite{kandel99,kandel95}
we obtain well decorated islands with impurities floating
on the island edges during growth (Fig.~\ref{fig:passivation}b, and c),
see Section~\ref{Morphology}. 

In principle, the rate of the exchange process could depend on
the number of nearest-neighbor bonds (zero or one) of the $A$-atom. 
We observed that the difference of both rates is not crucial
provided both processes are active.
The rates of these processes are taken as
$k_{\rm ex}=k_0 \exp (- E_{\rm ex}/k_B T)$,
where $E_{\rm ex}$ are the corresponding activation barriers.
For both processes there is a maximum activation barrier
above which the decorated geometry is not observed.
In the following, the barriers for 
both types of exchange are for simplicity set to be equal.

\section{Island morphology}
\label{Morphology}

\begin{figure}[hb]
\centering
\vspace*{100mm}
\includegraphics{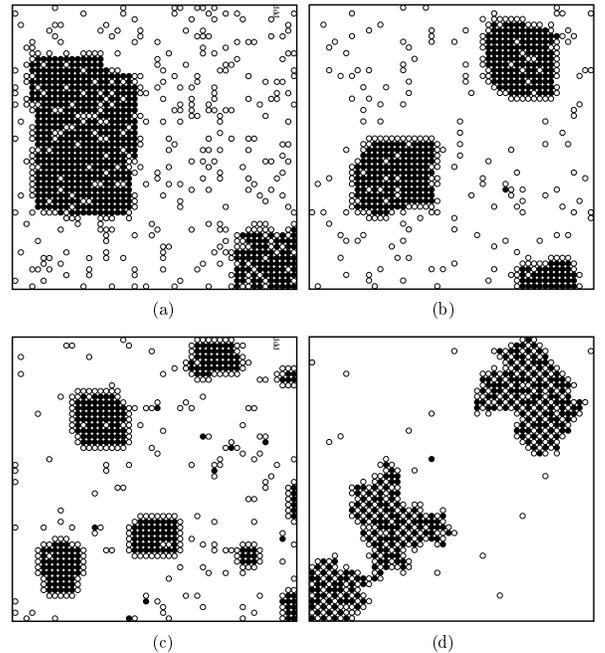}
\caption{Examples of configurations for the total coverage 
$\theta = 0.2$~ML
obtained by simultaneous deposition with flux $F_A=F_B = 0.004$~ML/s,
$E_{\rm n}^{AA} = 0.3$~eV and different energy barriers
$E_{\rm n}^{AB}$  and $E_{\rm ex}$:
(a)  $E_{\rm n}^{AB} = 0.1$~eV,   $E_{\rm ex}= 1$~eV, 
(b)  $E_{\rm n}^{AB} = 0.2$~eV,   $E_{\rm ex}= 1$~eV, 
(c)  $E_{\rm n}^{AB} = 0.4$~eV,   $E_{\rm ex}= 1$~eV, 
(d)  $E_{\rm n}^{AB} = 0.2$~eV,   $E_{\rm ex}= 2$~eV. 
We show only 50 $\times$ 50 sections of larger simulation
boxes.
}
\label{fig:passivation}
\end{figure}

In Fig. \ref{fig:passivation} we 
show examples of typical configurations with the same partial coverage 
of both species $\theta_A = \theta_B = 0.1$~ML (i.e., the total coverage
$\theta = \theta_A + \theta_B  = 0.2$~ML)
obtained by codeposition of adatoms and impurities with fluxes 
$F_A=F_B = 0.004$~ML/s.
Figures \ref{fig:passivation}a, \ref{fig:passivation}b, 
\ref{fig:passivation}c illustrate the effect of varying the relation between
$E_{\rm n}^{AA}$ and   $E_{\rm n}^{AB}$ for $E_{\rm n}^{AA} = 0.3$~eV.
Several features can be identified. First, the
island density increases with $E_{\rm n}^{AB}$.
This can be explained by the observation (cf. Fig.~\ref{fig:passivation}c  
for $E_{\rm n}^{AB}= 0.4$~eV) that
for larger $E_{\rm n}^{AB}$, free $A$-atoms start
to be captured by impurities
and many small islands containing
impurities and a few $A$ atoms appear on the surface
in addition to already existing decorated islands.
These small islands act as nucleation centers that lead
to the increase of the island density.
For large  $E_{\rm n}^{AB}$, almost all 
impurities will capture an $A$ adatom 
for a certain time. As we argue in Section~\ref{Rateq}, 
this effect causes reduction of adatom
mobility by the impurities. 

Whereas the island density increases, the
density of free impurities decreases with 
increasing $E_{\rm n}^{AB}$ due to
(i) the stronger $A$-$B$ bond favoring the
binding of impurities at island edges and (ii)
the increase of the island density that leads to
smaller islands with more perimeter sites. 
(Notice that we do not obtain decorated islands for simultaneous 
deposition at very large fluxes or at very early stages
of growth since there are not enough impurities available to 
cover all perimeter sites.)

The degree of edge decoration also strongly depends on the 
value of $E_{\rm n}^{AB}$.
Edge decoration is not observed for small $E_{\rm n}^{AB} = 0.1$~eV,
it is only partial for $E_{\rm n}^{AB} = 0.2$~eV, 
and it becomes  perfect for $E_{\rm n}^{AB}= 0.4$~eV
(cf. Figure~\ref{fig:passivation}).
Hence, in order to get decorated islands,
the barrier  $E_{\rm n}^{AB}$ has to be larger than a
minimal value. A simple detailed balance argument \cite{ecoss18}
shows that the fraction $f_0$ of uncovered edge sites is given by
\begin{equation}
\label{f0}
f_0 = (1 + \theta_B e^{E_{\rm n}^{AB}/k_B T})^{-1},
\end{equation}
and thus at $T = 500$~K a barrier of $E_{\rm n}^{AB} \geq 0.2$~eV is required.
As we shall see in Section~\ref{Rateq}, the condition $f_0 \ll 1$
also implies that the diffusion of $A$-atoms is slowed down considerably
by the impurities.

Another important parameter determining decoration of island edges
is the exchange barrier $E_{\rm ex}$ that was in our simulations
varied from 0.8~eV to 2~eV. 
For a small value of $E_{\rm ex}$, impurities are
driven toward island edges, 
whereas for large $E_{\rm ex}$, the  
exchange process is not active and
impurities are often incorporated inside the islands.
The impurities were observed to be floating on the island edges 
for small $E_{\rm ex}$, both for
$E_{\rm n}^{AA}  > E_{\rm n}^{AB}$ (Fig.~\ref{fig:passivation}b)
and for $E_{\rm n}^{AA}  < E_{\rm n}^{AB}$ (Fig.~\ref{fig:passivation}c).
Fig.~\ref{fig:passivation}d shows a configuration 
for $E_{\rm ex} = 2$~eV
in the case $E_{\rm n}^{AA} > E_{\rm n}^{AB}$.
The surface morphology is similar to the one observed for
$E_ {\rm n}^{AA} < E_{\rm n}^{AB}$ (not shown), where quite a regular
checkerboard 
\noindent
structure is produced with almost no 
free impurities on the surface.
Thus in order to obtain decorated islands, $E_{\rm ex}$ has to be lower
than a threshold value which in the present case is about 1.2~eV.

A further remarkable feature of the configurations displayed in 
Figure~\ref{fig:passivation}(b) and (c) 
is that the compact square island
shape is maintained as the island density increases. In fact,
careful inspection shows that the kink density on the well decorated
edges is {\em smaller\/} than when the decoration is incomplete.
This reflects the {\em enhancement of edge diffusion\/} by
the impurities: The energy barrier for an adatom
moving along the decorated step edge within the impurity layer 
is $E_{\rm ex}$ which, under the conditions of 
Figure~\ref{fig:passivation}(b),(c), is smaller than both
the barrier $E_{\rm sub}^{AA} + E_{\rm n}^{AA}$ for 
diffusion along an uncovered step and 
the barrier $E_{\rm sub}^{AA} + E_{\rm n}^{AB}$ for edge
diffusion on the outside of the impurity layer. 
Clearly this is true only if the exchange of singly bonded
$A$-atoms is allowed, which underlines the importance of
this type of the exchange process. 

\section{Island density scaling}
\label{Island}

\subsection{Simultaneous growth}

In the previous section, we described qualitatively the growth morphology 
for one fixed value of the deposition flux. 
Here we present results for the behavior of the island density
$N$ as a function of flux $F$ and coverage $\theta$, and discuss
its dependence on the kinetic parameters. 
Results for each set of parameters were obtained by averaging
over several independent simulation runs.

In the presence of impurities, islands are composed of both 
$A$ and $B$ atoms.
We define the size of an island as the number of $A$-atoms in a 
connected cluster of $A$-atoms forming the island.
This definition is appropriate for growth with impurities
segregating on the edges of the islands.
However, visual inspection of configurations showed that for
$E_{\rm ex} > 1.2$, there exist islands containing several mutually
disconnected clusters of $A$-atoms. Hence, our definition cannot
be applied straightforwardly for large $E_{\rm ex}$. 
In the following, we restrict ourselves to situations where the
intermixing inside the islands is negligible.
Simulations for larger values of $E_{\rm ex}$ indicate that
the flux dependence of the island density flattens
(the exponent $\chi$ in (\ref{N}) decreases), but 
due to the ambiguity in the definition of the island density in 
presence of intermixing, we did not attempt
to assess the physical significance of this observation.

\subsubsection{Flux dependence}
\label{SimFlux}

\begin{figure}[b]
\centering
\includegraphics{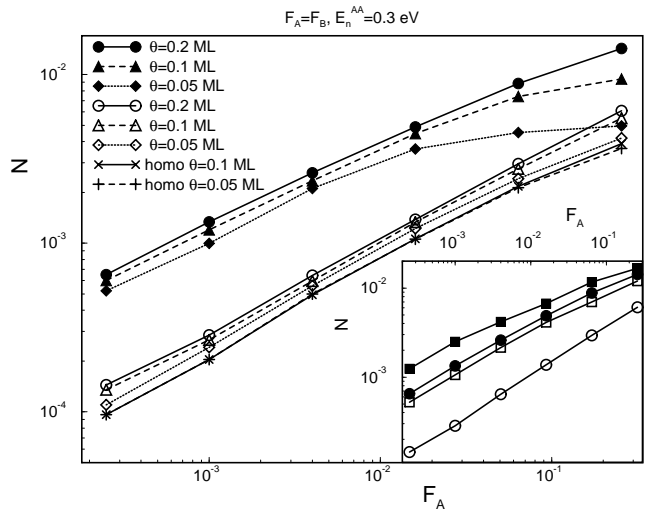}
\vspace*{73mm}
\caption{
Averaged island density as a function of flux
$F_A$ for several values of the total coverage 
$\theta = \theta_A + \theta_B$ and different  energy barriers:
$E_{\rm n}^{AB} = 0.2$~eV - open symbols,   
$E_{\rm n}^{AB} = 0.4$~eV - filled symbols.
The adatom interaction energy
$E_{\rm n}^{AA} = 0.3$~eV and  
the exchange barrier $E_{\rm ex}=1$~eV
are fixed, and the impurity flux $F_B = F_A$. 
The behavior in the absence of impurities
(homoepitaxy, $F_B = 0$) is shown for comparison.
Inset: The effect of the exchange barrier $E_{\rm ex}$ on 
the island density at coverage $\theta = 0.2$~ML.
Circles represent the data from the main figure for $E_{\rm ex} = 1$~eV,
squares the corresponding data for $E_{\rm ex} = 1.2$~eV.
}
\label{fig:fluxCod}
\end{figure}

Fig.~\ref{fig:fluxCod} shows the island density $N$ as a
function of the adatom flux $F_A$  for
several coverages $\theta$ and different energy barriers $E_{\rm n}^{AB}$ 
and $E_{\rm ex}$ (inset). 
The  energy barrier $E_{\rm n}^{AA} = 0.3$~eV  is
fixed, and the impurities and adatoms are codeposited with the same
flux $F_B = F_A$.
For comparison we also show data for homoepitaxial growth without
impurities at two coverages $\theta = 0.05$~ML and $\theta = 0.1$~ML. 
We can see that for $E_{\rm n}^{AB} = 0.2$~eV and $E_{\rm ex} = 1$~eV,
the island density is quite
close to the corresponding value in homoepitaxy. 
With increasing interaction energy between adatoms and impurities,
the island density dramatically increases, but the exponent $\chi$
in the power law relation (\ref{N}) between flux and the island density
remains nearly unchanged. For example, 
we find $\chi \approx 0.54$ for $E_{\rm n}^{AB} = 0.2$,
$\chi \approx 0.45$ for $E_{\rm n}^{AB} = 0.4$, and
$\chi \approx 0.54$ for homoepitaxial growth, which means that the 
effective critical nucleus size is $i^\ast \approx 2$ in 
this range of parameters. According to 
Kandel's rate equation theory,\cite{kandel97} the scaling
exponent should then become $\chi \approx 0.8$ in the presence
of strong barriers to attachment. In our model this is not
observed, because the bonding of the adatoms to the impurity-covered
edges keeps them near the edge long enough for an exchange to
occur.\cite{ecoss18,myslivecek00}

The inset of Fig.~\ref{fig:fluxCod} shows
that the island density is further increased if the 
exchange barrier $E_{\rm ex}$ is set to a larger value 
$E_{\rm ex} = 1.2$~eV. The data for $E_{\rm ex} = 1.2$~eV and
$E_{\rm n}^{AB} = 0.4$ indicate a slight decrease of the exponent
$\chi$. 

\begin{figure}[th]
\centering
\includegraphics{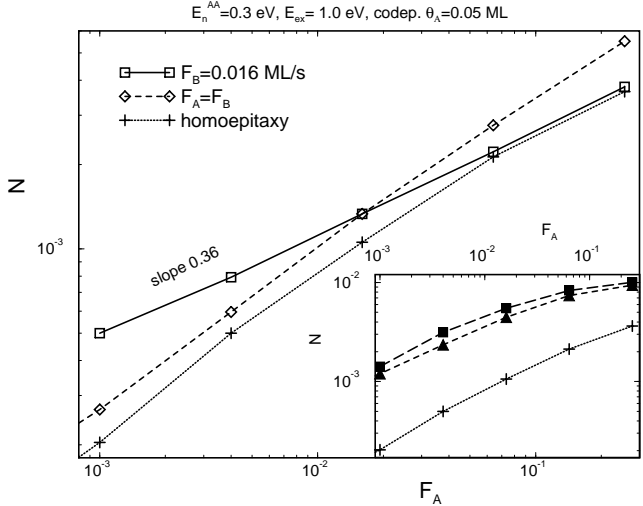}
\vspace*{65mm}
\caption{
Dependence of the island density on flux $F_A$
at constant  $F_B = 0.016$~ML/s
(open squares) compared with the situation  $F_B =  F_A$ 
(open diamonds), and with homoepitaxy (pluses), 
$E_{\rm n}^{AB} = 0.2$~eV. 
Inset: The flux dependence at two constant
ratios of fluxes, $F_B/F_A = 2$ (filled squares) and 
$F_B/F_A = 1$ (filled triangles), $E_{\rm n}^{AB} = 0.4$~eV  
compared with homoepitaxy (pluses).
}
\label{fig:constB}
\end{figure}

Figure~\ref{fig:constB} shows results obtained by varying the
ratio $F_B/F_A$ of impurity to adatom flux. In one set of
simulations, using $E_{\rm n}^{AB} = 0.2$~eV, 
the impurity flux was kept constant at $F_B = 0.016$~ML/s, 
while the adatom flux $F_A$ was varied. 
For large fluxes $F_A > F_B$, the island
density 

\noindent
is seen to approach the data obtained for homoepitaxy, 
indicating that the impurities have no effect, while for
small fluxes $F_A < F_B$, the flux dependence is described by an 
effective power law $N \sim F_A^{\chi'}$ with $\chi^\prime \approx 0.36 
< \chi$. An interpretation of this behavior will be given
at the end of Section~\ref{RateqCod}

In the second set of simulations shown in Figure~\ref{fig:constB},
which were carried out using $E_{\rm n}^{AB} = 0.4$~eV,
both fluxes were varied keeping the ratio $F_B/F_A = 2$ constant.
This is seen to further 
increase the island density without changing the flux dependence.
In this sense, an increase in the coverage of impurities (by increasing
$F_B$) is equivalent to increasing their effectiveness through an
increase of the bond energy $E_n^{AB}$. A quantitative formulation of
this statement will be given in Section~\ref{Rateq}.

\subsubsection{Coverage dependence}
\begin{figure}[th]
\centering
\includegraphics{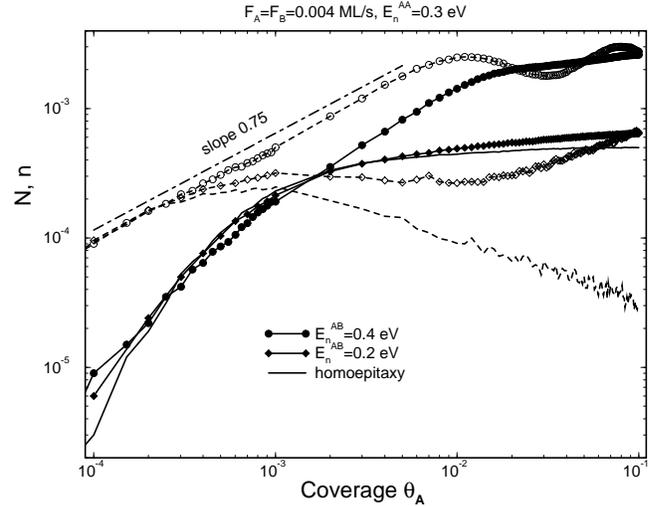}
\vspace*{65mm}
\caption{
Density of islands (solid lines) and free adatoms (dashed
lines) as a function of coverage $\theta_A$ for homoepitaxy 
(no symbols) and  for two different  energy barriers
$E_{\rm n}^{AB} = 0.2$~eV (diamonds) and
$E_{\rm n}^{AB} = 0.4$~eV (circles).
Adatoms and impurities are codeposited at the same flux,
$F_A = F_B = 0.004$~ML/s.
}
\label{fig:timeCo}
\end{figure}

A new feature in comparison with homoepitaxy is a stronger coverage
dependence of the island density. This is seen in Fig.~\ref{fig:fluxCod} 
for both weak ($E_{\rm n}^{AB} = 0.2$~eV)
and strong ($E_{\rm n}^{AB} = 0.4$~eV) 
interaction with impurities, but it is more 
pronounced for strong interaction, in particular at larger fluxes.
We followed the coverage dependence in more detail for fixed flux,
and in addition to the island density we also measured the density of free 
adatoms $n$. The results obtained at a medium flux $F_A =0.004$~ML/s
are compared with homoepitaxy in Fig.~\ref{fig:timeCo}.
Both in impure growth and in homoepitaxy the island density
shows an initial regime of rapid increase followed by a 
``saturation'' regime in which it increases much more slowly
with coverage. However, the residual coverage dependence in 
\noindent
the saturation regime is stronger in the presence of impurities, and
furthermore the onset of the saturation regime is delayed 
as the interaction between adatoms and impurities increases.
A quantitative description of
this effect will be provided in Section~\ref{RateqCod}.

The density of adatoms exhibits a completely different behavior as
compared to
homoepitaxy. We observe that for growth with impurities,
the adatom density is comparable with the island density 
up to the coverage $\theta_A = 0.1$~ML, whereas in homoepitaxy 
the adatom density rapidly decreases after reaching
a maximum at the beginning of the saturation regime 
(cf. Section~\ref{RateqPure}). 
Other surprising features are the power-law increase
$n \sim \theta_A^{0.75}$ observed over almost two decades
in the case of strongly interacting impurities, and the weak
oscillations of the adatom density for coverages $\theta_A > 0.01$
(cf. Fig.~\ref{fig:timeCo}). We will return to the behavior of the adatom
density in Section~\ref{RateqCod}.

\subsubsection{Next-nearest-neighbor interaction}

A modification of the model 
in which the barrier for diffusion $E^X_B$
contains an additional contribution from 
each lateral next-nearest neighbor of the
opposite type was also studied. This implies that a term
$n_2^{AB}E_{\rm nn}^{AB}$ is added to the right hand side of
Eq.~(\ref{E_D^X}).
Here $n_2^{AB}$ is the number of next-nearest neighbors 
of the type opposite to the atom under consideration,
and $E_{\rm nn}^{AB}$
is the corresponding contribution to the activation barrier.
We do not consider next-nearest-neighbor contributions
from pairs of particles of the same type.
For simplicity, the new parameter $E_{\rm nn}^{AB}$ is set 
equal to $E_{\rm n}^{AB}$.

\begin{figure}[hb]
\centering
\vspace*{55mm}
\includegraphics{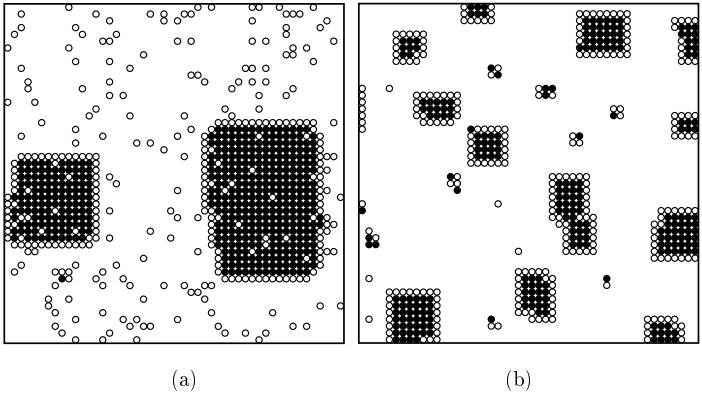}
\caption{Examples of configurations for the total coverage 
$\theta = 0.2$~ML
obtained by simultaneous deposition with flux $F_A=F_B = 0.004$~ML/s,
$E_{\rm n}^{AA} = 0.3$~eV, $E_{\rm ex}= 1$~eV
in a modified model with next-nearest-neighbor interaction:
(a)  $E_{\rm n}^{AB} = E_{\rm nn}^{AB} = 0.1$~eV,  
(b)  $E_{\rm n}^{AB} = E_{\rm nn}^{AB} = 0.4$~eV.
We show only $50 \times 50$ sections of larger simulation
boxes.
}
\label{fig:nnn}
\end{figure}
\noindent 

Our motivation for introducing the additional term is a desire
to study an improvement in the decoration of island
edges by impurities, and the resulting decrease of
the fraction $f_0$ of uncovered edge sites.
The additional interaction also enhances the
probability of nucleation around impurities
because the number of sites at which an adatom can be captured
is considerably higher.
The configuration shown in Fig.~\ref{fig:nnn}a demonstrates that
now we obtain almost perfect decoration also 
for  $E_{\rm n}^{AB}= E_{\rm nn}^{AB} = 0.1$~eV.
Fig.~\ref{fig:nnn}b illustrates the nucleation of small islands.

For  $E_{\rm n}^{AB}= E_{\rm nn}^{AB} = 0.1$~eV,
the island density is
nearly the same as for homoepitaxy for all fluxes studied,
and decoration is perfect provided there is sufficient amount
of impurities available. This shows that also in the presence 
of next-nearest-neighbor interactions the decorated
\begin{figure}[hb]
\centering
\vspace*{55mm}
\includegraphics{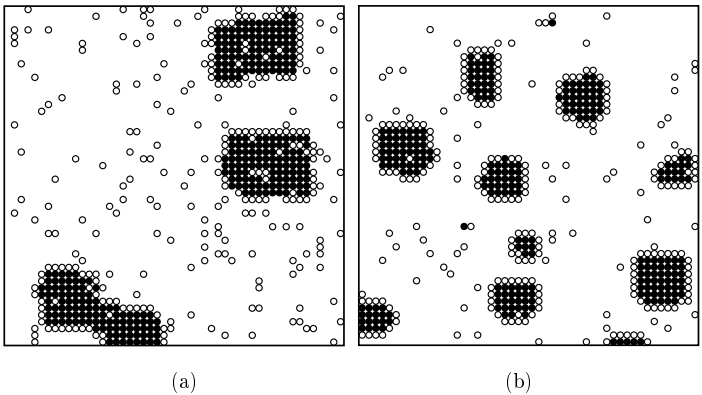}
\caption{Examples of configurations deposited with flux 
$F_A = 0.004$~ML/s
at the coverage $\theta_A = 0.1$~ML
after predeposition of $\theta_B = 0.1$, using parameters:
$E_{\rm n}^{AA} = 0.3$~eV, $E_{\rm ex}= 1$~eV,
(a)  $E_{\rm n}^{AB} = E_{\rm nn}^{AB} = 0.2$~eV,  
(b)  $E_{\rm n}^{AB} = E_{\rm nn}^{AB} = 0.4$~eV.
We show only $50 \times 50$ sections of larger simulation
boxes.
}
\label{fig:pre}
\end{figure}
\noindent
edges are 
unable to block efficiently the attachment of adatoms. 
For larger values of  $E_{\rm n}^{AB}= E_{\rm nn}^{AB}$,
the additional interaction causes an increase of the island density
and a {\em decrease\/} of the scaling exponent $\chi$. 
For example, the effective value 
for  $E_{\rm n}^{AB}= E_{\rm nn}^{AB} = 0.2$~eV is $\chi \approx 0.42$
and for  $E_{\rm n}^{AB}= E_{\rm nn}^{AB} = 0.4$~eV  it drops to
$\chi \approx 0.3$.
This suggests that the nucleation of small islands described above 
effectively lowers the size $i^\ast$ of the critical nucleus.

\subsection{Predeposition of impurities}

We performed simulations with predeposition of impurities
for the same set of parameters as for codeposition.
In order to obtain a morphology with island edges 
decorated by impurities, we need an appropriate 
value of $E_{\rm ex}$. Complete decoration also requires
a sufficient amount of impurities available on the surface.
The data presented are for a predeposited 
coverage $\theta_B =0.1$~ML.
Examples of morphologies for $E_{\rm n}^{AB} = 0.2$~eV and 
$E_{\rm n}^{AB} = 0.4$~eV  are shown in Fig.~\ref{fig:pre}
and look qualitatively similar to Figures~\ref{fig:passivation}
b and c.

\begin{figure}[hb]
\centering
\includegraphics{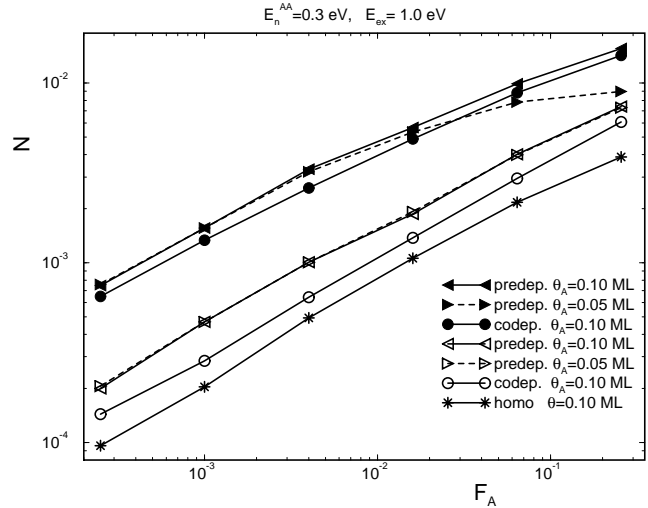}
\vspace*{70mm}
\caption{
Comparison of averaged island density as a function of flux $F_A$
for codeposition (circles)
and predeposition (triangles) for different  energy barriers:
$E_{\rm n}^{AB} = 0.2$~eV - open symbols,   
$E_{\rm n}^{AB} = 0.4$~eV - filled symbols.
The predeposited coverage is $\theta_B = 0.1$~ML.
The adatom interaction energy
$E_{\rm n}^{AA} = 0.3$~eV and  
the exchange barrier $E_{\rm ex}=1$~eV are fixed.
The behavior for homoepitaxy is shown for comparison.
}
\label{fig:pre-code}
\end{figure}

The $F_A$-dependence of the island density is compared with the results 
for codeposition in Fig.~\ref{fig:pre-code}.
The island densities in the predeposition regime are slightly higher
than for codeposition. This is qualitatively plausible, since the
predeposited impurities are present on the surface throughout the
deposition process and hence their effect on growth accumulates
over time. The corresponding curves are shifted by 
a factor independent of the flux. The slope remains the same as 
for codeposition.
The difference from codeposition is that there is no  appreciable

\begin{figure}[th]
\centering
\includegraphics{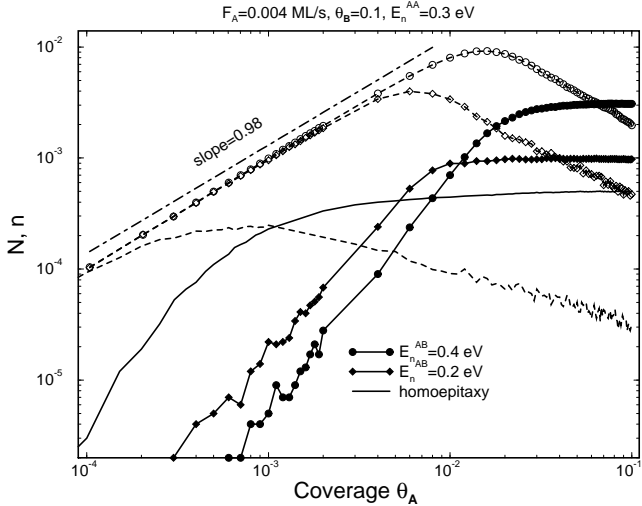}
\vspace*{70mm}
\caption{
Density of islands (solid lines) and free adatoms (dashed
lines) as a function of coverage $\theta_A$ for homoepitaxy 
(no symbols)
and for predeposition with two different  energy barriers 
$E_{\rm n}^{AB} = 0.2$~eV (diamonds) and
$E_{\rm n}^{AB} = 0.4$~eV (circles).
Flux of adatoms is $F_A = 0.004$~ML/s, the predeposited
coverage of impurities is $\theta_B = 0.1$~ML.
}
\label{fig:timePre}
\end{figure}
\noindent
coverage dependence 
for $\theta_A$ larger than $0.05$~ML, except at larger fluxes
in the case of strongly interacting impurities.

The detailed coverage dependence for a fixed flux is shown in 
Fig.~\ref{fig:timePre}. We see that the behavior of both
the island density and the adatom density is qualitatively
similar to homoepitaxy. The island density saturates 
and at the same time the adatom density starts to decrease.
The impurities only cause a shift 
of the crossover to the saturation regime
to a higher coverage, the shift being larger for 
stronger interaction between adatoms and impurities.
We shall return to this effect in Section~\ref{RateqPre}.

\section{Rate equation theory}
\label{Rateq}

In this section, we develop a simple rate equation approach to 
explain, at least qualitatively, the main impurity effects
on the island density which were presented in Section~\ref{Island}. 
Our basic assumption is that the impurities
affect the growth process {\em only by slowing down the
diffusion of adatoms\/}. 
To obtain a simple analytic expression
for the effective adatom diffusion coefficient
$\overline D(\theta_B)$ in the presence of an impurity coverage
$\theta_B$, we further replace the (mobile) impurities by
static traps with binding energy
$E_{\rm n}^{AB}$. 
Then standard results for diffusion in random
media yield \cite{haus82}
\begin{equation}
\label{Dbar}
\overline D(\theta_B) = \frac{D}{1 - \theta_B + 
\theta_B e^{E_{\rm n}^{AB}/k_B T}} 
\equiv \frac{D}{1 + \theta_B \phi} 
\end{equation}
where $D = k_0 e^{-E_{\rm sub}^{AA}/k_B T}$ is the diffusion
coefficient of a single adatom on the clean substrate,
and the abbreviation 
\begin{equation}
\label{phi}
\phi = e^{E_{\rm n}^{AB}/k_B T} - 1
\end{equation}
has been introduced.

The first conclusion that can be drawn from Eq.~(\ref{Dbar}) is that
predeposited impurities, $\theta_B$ = const., significantly affect 
the adatom diffusion only if 
$\theta_B \gg 1/\phi$.
In the case of codeposition $\theta_B = F_B t$ and 
${\overline D}$ becomes time- or coverage-dependent. It
is then useful to rewrite (\ref{Dbar}) in terms of the
coverage $\theta_A$ of $A$-atoms as
\begin{equation}
\label{Dbar2}
\overline D = \frac{D}{1 + 
\theta_A/\theta^\ast},
\end{equation}
with the characteristic coverage
\begin{equation}
\label{thetastar}
\theta^\ast = \frac{F_A}{F_B} \phi^{-1}.
\end{equation}
For coverages $\theta_A \geq \theta^\ast$ the impurities begin 
to significantly affect the adatom mobility.
The expression (\ref{thetastar}) quantifies the statement made
above in Section~\ref{SimFlux} that an increase of the flux
ratio $F_B/F_A$ in codeposition is equivalent to an increase of
the $A$-$B$ binding energy $E_{\rm n}^{AB}$. 
For $F_A = F_B$ and $T = 500$ K, we have $\theta^\ast \approx 0.01$ for
$E_{\rm n}^{AB} = 0.2$~eV and $\theta^\ast \approx 10^{-4}$ for 
$E_{\rm n}^{AB} = 0.4$~eV. In the following, these two sets of 
parameters will be referred to as the case of weak and strong
impurities, respectively.

\subsection{Pure growth}
\label{RateqPure}

We proceed by combining (\ref{Dbar}) with the simplest
analytic model of nucleation, consisting of two coupled rate
equations for the island density $N$ and the adatom density
$n$. In the absence of impurities, the equations 
for a critical island size $i^\ast = 1$
read \cite{stoyanov81,venables84,tang93}
\begin{equation}
\label{raten}
\frac{dn}{dt} = F_A - 4 D n( 2n + N) 
\end{equation}
\begin{equation}
\label{rateN}
\frac{dN}{dt} = 4 D n^2.
\end{equation}
 
The main features of 
the solution of (\ref{raten},\ref{rateN})
with initial condition $n = N = 0$ can be described as follows
(see the article by Tang\cite{tang93} 
for a lucid presentation): In the 
{\em early time regime\/} the adatom density increases linearly
by deposition,
$n \approx F_A t = \theta_A$, and accordingly the island density
grows as 
\begin{equation}
\label{Nearly}
N \approx (4/3)(D/F_A)\theta_A^3.
\end{equation} 
In the 
{\em late time regime\/} the adatoms are mainly captured by
preexisting islands. This implies that $n \approx F_A/D N \ll N$
and the island density grows more slowly, as 
\begin{equation}
\label{Nlate}
N \approx (F_A/12 D)^{1/3} \theta_A^{1/3},
\end{equation}
while the adatom density decreases as $n \sim \theta_A^{-1/3}$. 
The transition between the two regimes 
occurs at a coverage
\begin{equation}
\label{theta1}
\theta_1 \sim (F_A/D)^{1/2}.
\end{equation}
Keeping the coverage fixed while increasing the flux therefore
takes the system from the late time regime, where $N \sim F^{1/3}$,
into the early time regime with $N \sim F^{-1}$, with a maximum
in the island density attained at a critical flux
$F^c \sim D \theta^2$.

To generalize these estimates to the case $i^\ast > 1$, we replace
the nucleation equation (\ref{rateN}) by \cite{stoyanov81,venables84}
\begin{equation}
\label{rateNgen}
\frac{dN}{dt} \sim D n^{i^\ast + 1}.
\end{equation}
Then the early time behavior becomes $N \sim (D/F) \theta_A^{i^\ast
+ 2}$, while in the late time regime
\begin{equation}
\label{Nlategen}
N \sim (F_A/D)^{i^\ast/(i^\ast + 2)} \theta_A^{1/(i^\ast + 2)},
\end{equation}
in agreement with the expression (\ref{gamma}) for the scaling
exponent $\chi$. 
The transition coverage $\theta_1$ is estimated by matching the
two behaviors, which yields
\begin{equation}
\label{theta1gen}
\theta_1 \sim (F_A/D)^{2/(i^\ast + 3)}.
\end{equation}

For $i^\ast = 1$ the coverage dependence of the densities of islands
and adatoms observed in microscopic simulations is in accordance
with the rate equation theory.\cite{tang93,evans93} 
In the reversible case $i^\ast > 1$,
the simple rate equations are quantitatively
inappropriate, though the key qualitative features -- the
existence of an early time regime of a rapid increase of the
island density, followed by a ``precoalescence saturation regime''
with little change in $N$ -- remain.\cite{ratsch94}  

\subsection{Impure growth}

\subsubsection{Predeposition}
\label{RateqPre}

The effect of predeposited impurities is obtained simply 
by replacing $D$ by the constant expression
(\ref{Dbar}) for ${\overline D}$ 
in (\ref{Nlategen}) and (\ref{theta1gen}).
Consequently the island density in the late
time regime increases by a factor $(1 + \theta_B \phi)^{i^\ast/
(i^\ast + 2)}$ which is
independent of flux or $A$-coverage, and the onset
of saturation is delayed 
by a factor $(1 + \theta_B \phi)^{2/(i^\ast + 3)}$.
This is in qualitative agreement with the coverage dependence 
of densities of islands and adatoms displayed in 
Figure~\ref{fig:timePre},
which shows the same overall behavior as in the homoepitaxial case, only
shifted to larger coverages and higher densities. Quantitatively,
the numerically observed increase in the island density is consistent
with the factor $(1 + \theta_B \phi)^{i^\ast/(i^\ast + 2)}$
if the size of the critical nucleus is set to $i^\ast = 1$. On the
other hand, if the critical nucleus size is
assumed to be $i^\ast = 2$ as suggested by the numerical value of 
$\chi$, then the theory is seen to overestimate the increase in the
island density and underestimate the delay of the onset of 
saturation. 

As was mentioned in Section~\ref{RateqPure}, the island density at fixed
coverage $\theta_A$
shows a maximum at a critical flux $F^c$, which is determined
by setting the saturation coverage equal to $\theta_A$. For
predeposited impurities with $\theta_B \phi \gg 1$ this is given by
\begin{equation}
\label{Fcpredep}
F_A^c \sim (D/\phi \theta_B) \theta_A^{(i^\ast + 3)/2}.
\end{equation} 
Further discussion of predeposition in relation to 
codeposition will be provided below.

\subsubsection{Codeposition}
\label{RateqCod}

In the
case of codeposition the situation is richer due to the 
coverage dependence of ${\overline D}$. First, the
cases $\theta_1 < \theta^\ast$ and $\theta_1 > \theta^\ast$
have to be distinguished. In the first case the impurities only
affect the late time regime. For $\theta_A \gg \theta^\ast$
we can approximate (\ref{Dbar2}) by ${\overline D} \approx
D \theta^\ast/\theta_A$. Inserting this into the nucleation 
equation (\ref{rateNgen}) and setting $n \approx F/ {\overline D}
N$ we obtain the expression
\begin{equation}
\label{Nimpur}
N \approx (F_A/D \theta^\ast)^{i^\ast/(i^\ast + 2)} 
(\theta_A)^{(i^\ast + 1)/(i^\ast + 2)}
\end{equation}
which replaces (\ref{Nlategen}) in the pure case. 
It can be seen that the scaling of $N$ with flux $F_A$ remains the
same, i.e. the exponent $\chi$ is not affected.
The impurities increase
the island density by a factor $(\theta_A/\theta^\ast)^{i^\ast/(
i^\ast + 2)}$ which, in contrast to the case of predeposition,
is coverage dependent. This is qualitatively consistent with
the residual coverage dependence of the island density seen
in Figure~\ref{fig:fluxCod}, which is nearly absent in the corresponding
predeposition data in Figure~\ref{fig:pre-code}. In quantitative
terms, however, the coverage dependence in (\ref{Nimpur}) is much
stronger than what is observed in the simulations. The dependence
of (\ref{Nimpur})
on $\theta^\ast$ shows that increasing the ratio $F_B/F_A$ at
constant $F_A$ and $\theta_A$ will also increase the island density,
in accordance with the simulation data shown in Figure~\ref{fig:constB}. 

Noting that for codeposition $\theta_B = (F_B/F_A) \theta_A = 
\theta_A/(\phi \theta^\ast)$, Eq.~(\ref{Nimpur}) can be
rewritten as $N \approx (F_A \theta_B \phi/D)^{i^\ast/(i^\ast + 2)}
\theta_A^{1/(i^\ast + 2)}$, which is identical to the expression
for predeposition with $\theta_ B \phi \gg 1$. A more careful
calculation shows that the island density for predeposition
exceeds that for codeposition by a constant factor 
$(i^\ast + 1)^{1/(i\ast + 2)}$, if systems of the same impurity
coverage $\theta_B$ are compared. This is qualitatively consistent
with Figure~\ref{fig:pre-code}, though the simulation data indicate
that the factor is larger for weak impurities than for strong ones.

A surprising consequence of the rate equations with coverage-dependent
diffusion is that the adatom density {\em increases\/} with coverage in
the late time regime. This follows from balancing the deposition 
term on the right hand side of (\ref{raten}) against the island
capture term $4 {\overline D} N n \approx 4 D (\theta^\ast / \theta_A)
N n$, which yields 
\begin{equation}
\label{nimpur}
n \sim (F_A/D \theta^\ast)^{2/(i^\ast + 2)} \theta_A^{1/(i^\ast + 2)}.
\end{equation}
In fact, when $\theta_1 < \theta^\ast$ the adatom density
is a nonmonotonic function of coverage: It increases as 
$n \sim \theta_A$ for $\theta_A < \theta_1$, decreases as
$n \sim \theta_A^{-1/(i^\ast +2)}$ for $\theta_1 < \theta_A < \theta^\ast$,
\begin{figure}[hb]
\centering
\includegraphics{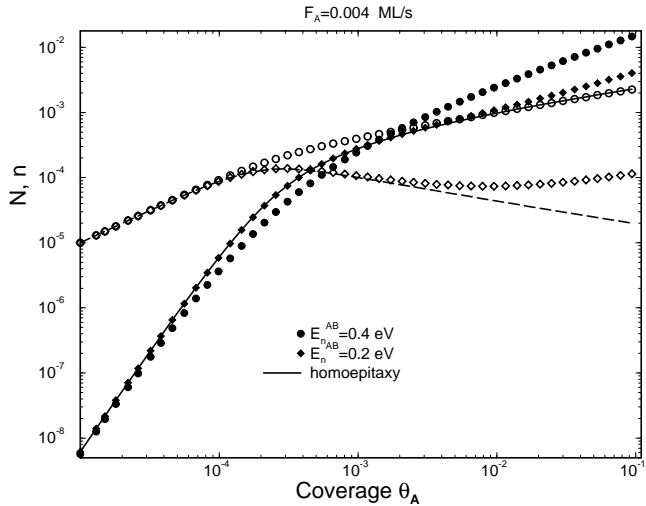}
\vspace*{70mm}
\caption{
Time evolution of the adatom (open symbols and the dashed line)
and island (full symbols and the solid line) densities obtained from
a numerical solution of the rate equations for
$i^\ast = 1$ {[}Eqs.~(\ref{raten}) and (\ref{rateN}){]}
with the coverage-dependent diffusion constant (\ref{Dbar}). The
parameters were chosen for comparison with the simulation data in 
Figure~\protect\ref{fig:timeCo}: Flux $F_A = F_B = 0.004$~ML/s, 
$D = (k_0/4) \exp (-E_{\rm sub}^{AA} /k_B T) \approx 
2.15 \times 10^4 {\rm s}^{-1}$, and $E_{\rm n}^{AB} = 0.2$~eV
(diamonds) and $E_{\rm n}^{AB} = 0.4$~eV (circles).
The agreement between the homoepitaxial island density (solid line) and
the adatom density for weak impurities (open circles) is coincidental.
}
\label{fig:timeRate}
\end{figure}
\noindent
and increases again according to (\ref{nimpur}) for $\theta_A > \theta^\ast$.
This is illustrated in Figure~\ref{fig:timeRate}, which is reminiscent of
the simulation data for the adatom density in Figure~\ref{fig:timeCo}.
However, in contrast to the simulations, the 
late time adatom density given by
(\ref{nimpur}) is small compared to the island density (\ref{Nimpur}),
and neither the intermediate scaling regime $n \sim (\theta_A)^{0.75}$
nor the oscillations of $n$ seen in Figure~\ref{fig:timeCo} are
reproduced by the rate equations. Here and in the following figures we
show results obtained by numerical integration of the rate equations
(\ref{raten},\ref{rateN}) for $i^\ast = 1$, with $D$ replaced
by $\overline D$. This is sufficient for a qualitative comparison,
and relieves us of the necessity to explicitly treat the growth
dynamics of the intermediate unstable clusters.

Consider next the case $\theta_1 > \theta^\ast$. For coverages
in the early time regime which satisfy $\theta_A \gg \theta^\ast$,
we set ${\overline D} \approx D \theta^\ast/\theta_A$ and obtain,
using $n \approx \theta_A$, the early time behavior
$N \sim (D \theta^\ast/F_A) \theta_A^{i^\ast+1}$. 
The onset of the saturation regime then occurs at a coverage
\begin{equation}
\label{theta1impur}
\tilde \theta_1 \sim (F_A/ D \theta^\ast)^{2/(i^\ast+1)}
\end{equation}
which exceeds the corresponding expression~(\ref{theta1gen}) for the
pure system by a factor $(\theta_1/\theta_\ast)^{2/(i^\ast + 1)}$.
Thus both predeposited and codeposited impurities delay the 
onset of the saturation regime. 

The critical flux at which the island density attains a maximum
is now given by
\begin{figure}[hb]
\centering
\includegraphics{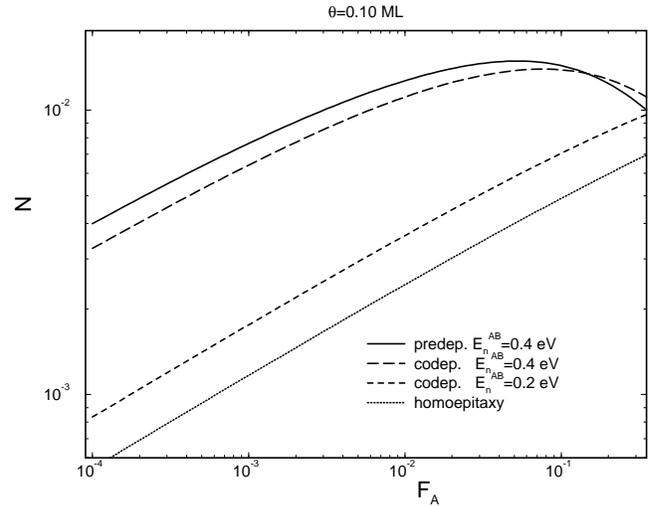}
\vspace*{70mm}
\caption{
Rate equation predictions for the flux dependence of the island density
in the case of codeposition with $F_A = F_B$ at the total coverage 
$\theta = 0.1$~ML/s. The parameters were chosen as in 
Figure~\ref{fig:timeRate}. 
For comparison with Figure~\ref{fig:pre}, data for predeposition with 
$\theta_B = \theta_A = 0.05$~ML/s in the case of strong impurities are shown 
as well. 
}
\label{fig:fluxRate}
\end{figure}

\begin{equation}
\label{Fccodep}
F_B^c \sim (D/\phi) \theta_A^{(i^\ast + 1)/2}
\end{equation}
which is seen to become identical to the predeposition expression
(\ref{Fcpredep}) by setting $\theta_B = (F_B/F_A) \theta_A$. 
A quantitative evaluation using the parameters of the simulations
\begin{figure}[hb]
\centering
\includegraphics{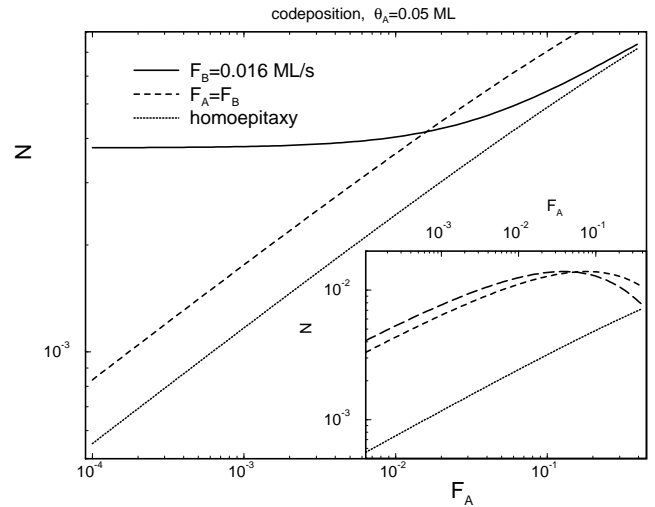}
\vspace*{70mm}
\caption{
This figure illustrates the 
rate equation prediction for codeposition 
with constant impurity flux $F_B = 0.016$~ML/s (solid line)
and should be compared to Figure~\ref{fig:constB}.
The predictions for the case $F_A = F_B$ (dashed line)
and for homoexpitaxy (dotted line) are also shown. 
$E_{\rm n}^{AB} = 0.2$~eV, 
other parameters were chosen as in Figure~\ref{fig:timeRate}. 
Inset: The rate equation prediction for  codeposition 
with constant flux ratio $F_B/F_A = 2$ (long-dashed line) 
compared to  the case $F_A = F_B$ (dashed line)
and to homoexpitaxy (dotted line),  $E_{\rm n}^{AB} = 0.4$~eV.
}
\label{fig:FBRate}
\end{figure}
\noindent
shows that the critical flux is beyond the range of simulated fluxes
for the case of weak impurities, while it
 should be observable
for strong impurities. This is illustrated in Figure~\ref{fig:fluxRate}.
It therefore appears natural to identify the saturation of the flux dependence
of the island density found in the simulations (see Figures~\ref{fig:fluxCod} 
and \ref{fig:pre-code}) with the maximum predicted
by the rate equations, although it should be emphasized that the simulations
show no clear evidence of a decrease of $N$ beyond the plateau.

Finally, we address the effect of changing the adatom flux
$F_A$ while keeping the impurity flux $F_B$ constant. At fixed 
$\theta_A$ this implies a crossover from $\theta_A \ll \theta^\ast$
for $F_A \gg F_B \phi \theta_A$ to $\theta_A \gg \theta^\ast$
for $F_A \ll F_B \phi \theta_A$. In the high flux regime the
island density $N $ is unaffected by the impurities, while in the
low flux regime $N \sim (F_A/D)^\chi (\theta_A/\theta^\ast)^\chi \sim 
(F_B \phi/D)^\chi$
becomes {\em independent} of $F_A$, since
$\theta^\ast \sim F_A$. This is illustrated in Figure~\ref{fig:FBRate},
which should be compared to Figure~\ref{fig:constB}. The behavior
is qualitatively similar, however instead of a plateau at low fluxes
the MC simulations show a second scaling regime $N \sim F_A^{\chi'}$ 
with a nonzero exponent $\chi' < \chi$.

\subsubsection{The effect of edge decoration}

In the formulation of the rate equations we have assumed that all
deposited impurities contribute to the impurity coverage
in the expression (\ref{Dbar}) for the effective diffusion coefficient,
thus neglecting the fact that a certain fraction of impurities is
bound at the island edges. This assumption is self-consistent only
if the density $n_e$ of edge sites, as predicted by the rate equations, is 
small compared to the deposited coverage $\theta_B$ 
of impurities at all times.

For compact islands the density of edge sites is of the
order of $N \sqrt{A}$, where $A \sim (\theta_A - n)/N$ is the area
of an island, and hence
\begin{equation}
\label{ne}
n_e \sim \sqrt{N(\theta_A - n)}.
\end{equation}
In the saturation regime $n, \; N \ll \theta_A$, thus 
$n_e \sim \sqrt{\theta_A N}$. In the early time regime
$n \approx \theta_A$ to leading order, and the density of edge sites
is determined by the next-to-leading correction. Analysis of the rate
equations shows that $n_e \sim N$ both for $\theta_A < 
\theta^\ast$ and for $\theta_A > \theta^\ast$, which implies that the
island size does not increase with coverage in this regime.

Using the estimates for $n_e$, the importance of the impurities bound
at the edges can be worked out for specific cases. 
Since evidently $n_e \ll \theta_A$ always, a {\em sufficient\/} condition for 
the irrelevance of edge decoration is that $\theta_B > \theta_A$ at
all times. This is true for predeposited impurities up to a coverage
$\theta_A  = \theta_B$, and for codeposition with $F_A \leq F_B$.
Among the situations treated earlier in this section, the only case 
where corrections due to edge decoration may be expected is deposition
at fixed impurity flux $F_B$ in the transition region $F_B \leq F_A \leq   
F_B \phi \theta_A$ (see Figure~\ref{fig:FBRate}). For fluxes 
$F_A > F_B \phi \theta_A$ the impurities were seen to be irrelevant
even if the full impurity coverage contributes to slowing down the
adatoms. Since the edge decoration decreases the impurity concentration
on the terraces, it is likely that its only effect will be to shift
the point where the island density becomes equal to its homoepitaxial
value towards smaller deposition fluxes.

\section{Conclusions}
\label{Conclusions} 

A brief glance at the results presented in this paper 
may lead to disappointment:
No dramatic change of the exponent $\chi$ has been found contrary
to expectations in any of the modifications of the simulation
model.\cite{myslivecek00}
However, upon closer inspection our work reveals several
nontrivial features that should not be overlooked in the large
amount of numerical data. 

(i) We have established
that a perfect decoration of island edges in our model requires 
a process of adatom exchange that is completely analogous to the
exchange discussed for surfactant-mediated growth of 
semiconductors.\cite{kandel95} 
A feature worth remembering is the {\em enhancement
of diffusion of adatoms along island edges\/} via the exchange process
with impurities attached to these edges. This mechanism of smoothing
on a one-dimensional substrate provides another perspective and 
possible interpretation of the smooth growth on a two-dimensional 
substrate in the presence of a surfactant. Smoothing of
island shapes should be experimentally verifiable  and
may have practical implications. 

Given the fact that size and shape of islands (including the number
of kinks at island edges) in the submonolayer regime of growth
determine the developing surface morphology (cf. most of the 
experimental papers cited below, and in particular 
Ref.~\onlinecite{kalff99}), 
a possibility to control {\em both\/} of them by adding 
impurities seems very attractive.

(ii) From a theoretical perspective, we have been able to obtain insight
into (and even semi-quantitative agreement with)
the simulation results, using rather simple rate equation
theory. The surprising resistance to change of the exponent $\chi$
can be understood within this theory, as well as other
features of the simulations, such as the strikingly different behavior
of the adatom density during impure growth with codeposition
as compared to the case of homoepitaxy.

Our research also leads to new questions. One would like to
understand better the details of behavior observed, in particular
the oscillations of the adatom density seen in simulations
(Fig.~\ref{fig:timePre}).
Another open question concerns ``monovalent'' impurities that can bond
to island edges but do not bond to adatoms approaching these
edges, and which therefore should bring about an efficient
passivation of the island boundaries.
Preliminary simulations suggest that this effect
alone does not bring about any significant change in the
value of the exponent $\chi$.

\vspace*{0.5cm}
\noindent
{\bf Acknowledgements.} JK acknowledges the kind hospitality of the
Institute of Physics, Czech Academy of Sciences, Prague, 
while part of this paper was prepared. 
This work was supported by Volkswagenstiftung, 
by the COST project P3.130 and by DFG within SFB 237.

\end{document}